\documentclass[epj]{svjour}

\usepackage{graphicx}
\usepackage{bm}
\usepackage{color}
\usepackage{amsmath}

\newcommand{\bsigma}{\mbox{\protect\boldmath $\sigma$}}
\begin{document}

\title{Dynamics and Performance of Susceptibility Propagation on Synthetic Data}

\author{Erik Aurell\inst{1,2,3} \and Charles Ollion \inst{1,3}\and Yasser Roudi\inst{4}}
\institute{ACCESS Linnaeus Center KTH-Royal Institute of Technology, 100 44 Stockholm, Sweden
\and Department of Informatics and Computer Science, Aalto University, Finland
\and Department of Computational Biology, AlbaNova University Centre, 106 91 Stockholm, Sweden
\and NORDITA, Roslagstullsbacken 23, 10691 Stockholm, Sweden}

\abstract{
We study the performance and convergence properties of the Susceptibility 
Propagation (SusP) algorithm for solving the Inverse Ising problem. We first study how the
temperature parameter ($T$) in a Sherrington-Kirkpatrick model generating the data influences the performance and
convergence of the algorithm. We find that at the high temperature regime ($T>4$), 
the algorithm performs well and its quality
is only limited by the quality of the supplied data. In the low temperature
regime ($T<4$), we find that the algorithm typically does not converge, yielding diverging
values for the couplings. However, we show that by stopping the algorithm 
at the right time before divergence becomes serious, good reconstruction can be
achieved down to $T\approx 2$. We then show that dense connectivity, loopiness of the connectivity,
and high absolute magnetization all have deteriorating effects on the performance of the
algorithm. When absolute magnetization is high, we show that other methods can be
work better than SusP. Finally, we show that for neural data with high 
absolute magnetization, SusP performs less well than TAP inversion.
}

\maketitle

\section{Introduction}
Problems in Statistical Mechanics and those in Statistical Inference 
can be thought of as being the inverses of each other. In statistical mechanics
one is usually given a Gibbs distribution and is asked to compute moments
of some observables. In statistical inference, one is given a set of observables
and is asked to reconstruct the distribution that generated them. Although the 
two fields have traditionally been developed separately, recently the 
connections and similarities have been highlighted, see \textit{e.g.} \cite{MezardMontanari09}.

A paradigmatic scenario in this direction is the Inverse Ising problem, 
that is finding the couplings of an Ising model given data \textit{e.g.} mean magnetizations 
and pairwise correlations. Two strands of questions in biology have recently 
motivated this problem. 
The first one comes from Neuroscience. While traditionally
it has only been possible to record 
simultaneously from a few neurons at a time, for special cases, \textit{e.g.} 
retinal cells recordings from hundreds of neurons are now possible,
and techniques allowing for many more are on the horizon. The problem is
then to infer functional connectivities between neurons from these recorded 
multi-neuronal spike trains.
\cite{Schneidman05,CoccoLeiblerMonasson,Roudi09-2,Roudi09-3}. 
Second, global gene expression measurements by \textit{e.g.} microarray 
technologies have been around for more than a decade, and a standard way 
to analyse them is through co-expression, \textit{i.e.} correlation of 
expression of genes or groups of genes across different conditions. 
Correlation is not causation. If, in fact, the gene expression measurements
are snap-shots of a probability distribution generated by an Ising model, 
then the most significantly coupled genes will be the ones most strongly 
coupled in the Ising model, not the most strongly
correlated, and this is then another way to classify genes as similarly or not similarly
expressed~\cite{Lezonetal06}. 

The inverse Ising problem is a difficult combinatorial
optimization problem in the class known as ``NP-hard''.
In theory, only approximate schemes, or methods that take more than polynomial time
to find the answer are possible. Boltzmann Learning \cite{Ackley85} is an
iterative method where in one step the 
correlation functions are computed given an Ising model, and in another step the Ising model couplings are
modified to adjust to data. In principle, Boltzmann learning can be employed to find the
couplings with arbitrary accuracy given accurate data and sufficient time, 
but the slow convergence of the Boltzmann learning makes it a very inefficient 
algorithm for most practical purposes. The main approximate schemes using
means and correlations are inversion
of the correlation matrix (``naive mean-field theory'') as used in~\cite{Lezonetal06},
Thouless-Anderson-Palmer (TAP) formula ~\cite{Kappen98,Tanaka98},
Independent Pairs Approximation \cite{Roudi09-2} and the perturbative scheme
of an auxiliary statistical mechanics model of Sessak and Monasson~\cite{SessakMonasson}.
If the sample data can be used as such (and not only via means and pairwise correlations), 
a linear regression can be performed on a link-by-link basis, which is quite powerful 
when the underlying matrix of couplings in sparse~\cite{RavikumarWainwrightLafferty10}. 

In this paper, we study the recently introduced Susceptibility Propagation (SusP), an approximate
scheme also based on means and correlations~\cite{MezardMora}. SusP is a message-passing algorithm 
which is derived by using the Fluctuation-Response theorem to the update 
rules of Belief Propagation. Although there are particular features specifically
for Ising spins, on one level, and more generally,
SusP can be considered as Boltzmann Learning scheme, where the 
correlation functions are computed by Belief Propagation
instead of (much more slowly) by Monte Carlo. 
Belief Propagation is based on messages exchanged between each pair of nodes \cite{YedidiaFreemanWeiss03}.
SusP uses such messages as well as the derivatives of these messages with 
respect to the field at a third node, yielding
in the end messages involving triplets of nodes. 
SusP is fairly sensitive to the accuracy by which the
correlation functions are known~\cite{Marinari10,OllionMScthesis}. 

Our contributions in this paper are as follows. 
In the paradigmatic Sherrington-Kirkpatrick model, also
studied in~\cite{MezardMora} and~\cite{Marinari10}, we provide
numerical evidence for a phase transition between \textit{reconstructible}
and \textit{non-reconstructible} phase, relative to the SusP algorithm. 
Perhaps surprisingly, this transition is not at the spin glass
transition ($T_c=1$) but some distance into the disordered paramagnetic phase 
($T_{SusP}\approx 4$).
We show that in the non-reconstructible but still disordered phase,
SusP almost converges, in the sense that trajectories of the updates
according to SusP come close to
a marginally unstable fixed point, and spend a long time in the neighborhood
of that fixed point before eventually diverging. We introduce a heuristic
stopping criterium for SusP in this region, and with a reasonable
criterion for quantitative reconstruction we are able to push the
threshold for (approximate) reconstruction down to $T_{SusP'}\approx 2$).
We also investigate how the performance of SusP
on a Sherrington-Kirkpatrick model depends on external field (magnetization).
Going beyond the Sherrington-Krirkpatrick model, we considered various
sparse models, where many (or most) of the potential couplings are zero.
In particular, we show that SusP works very well on a randomly diluted 
Sherrington-Kirkpatrick model, and in this scenario clearly outperforms
other approximate schemes. Furthermore, we show that, given a fixed sparsity,
the reconstruction is better on a randomly connected graph 
compared to a regular lattice. Finally, we show that
for \textit{in silico} neural data for which TAP inversion 
and Sessak-Monasson approximations work well ~\cite{Roudi09-2,Roudi09-3}, SusP 
appears to be out-performed by these two simpler schemes.

\section{General Setting}
\label{s:general-setting}
Susceptibility propagation is a fairly complicated algorithm,
which, for completeness, we describe in Appendix. 
To be able to refer to a definite formula, let us however state
that one central step in SusP is the inversion

\begin{equation}
\label{eq:SusP}
J_{ij} = \tanh^{-1}\left(\frac{C_{ij} - \tanh h_{i \rightarrow j} \tanh h_{j \rightarrow i}}{1 - C_{ij} \tanh h_{i \rightarrow j} \tanh h_{j \rightarrow i}}\right)
\end{equation}
where $h_{i \rightarrow j}$ and $h_{j \rightarrow i}$ are Belief Propagation messages
and $C_{ij}$ are auxilary quantities encoding gradients of messages,
the observed correlation between spins $i$ and $j$, as well as the observed magnetizations of
spin $i$ and spin $j$. After an update of the $J_{ij}$
the messages and their gradients are updated according to equations given in Appendix, where the
$J_{ij}$ enter parametrically, and this procedure is repeated until convergence.
When this procedure converges,
and how accurate it is when it does are important points to understand here.
In this Section we describe how we adress these two points
in this paper. 

The overall idea is to compute
mean magnetizations and pairwise correlations from
a \textit{known} Ising model and use them to 
reconstruct the model back. We can then illustrate
reconstruction by a scatter-plot of the inferred
couplings $J^{\rm SusP}_{ij}$ \textit{versus} the (known)
true couplings couplings $J_{ij}$. A straight line 
scatter plot of slope one indicates a successful reconstruction.
This can be quantified by the correlation coefficient of the
scatter plot ($R$),
and the P-value of a the null hypothesis that the reconstructed
couplings are equal to the true ones.
Alternatively, we can quantify the quality of reconstruction
in one of our test cases by the $\Delta$-measure of~\cite{MezardMora,Marinari10}:
\begin{equation}
\label{eq:SK}
\Delta=\frac{1}{{\rm std}(J_{ij})}\sqrt{\frac{2}{N (N-1)}\sum_{i<j} [J^{\rm SusP}_{ij}-J_{ij}]^2}
\end{equation}
The first test case is the Sherrington-Kirkpatrick 
model \cite{Sherrington75} of $N$ spins, $\bsigma=(\sigma_1,\dots,\sigma_N)$
and Boltzmann distribution
\begin{equation}
\Pr(\bsigma)=\frac{1}{Z} \exp\left[\beta \sum_{i} h_i \sigma_i+ \beta \sum_{i<j} J_{ij} \sigma_i \sigma_j\right],
\end{equation}

\noindent in which $\beta=1/T$ is the inverse temperature 
and the couplings $J_{ij}$ are drawn from a zero mean Gaussian distribution 
with variance $1/N$. This case has already been considered by~\cite{MezardMora},
where it was shown that SusP outperforms several other reconstruction methods
in the high temperature and zero external field regime (small $\beta$, all $h_i$ zero), 
and also by~\cite{Marinari10}
where it was shown that SusP is sensitive to noise in the correlation functions. 
Our contribution, for this family of test cases,
is a precise determination of the threshold in the
low temperature regime where SusP ceases to converge (in Section~\ref{s:convergence}),
and an extension of the analysis to the case with non-zero external fields (in Section~\ref{s:magnetization}).
We further introduce a modification of the SusP
stopping criterion pushing the boundary of (approximate) reconstruction to
lower temperatures. 

Motivated by the ``folklore theorem''
that message passing techniques work best on locally tree-like graphs, we also 
study the effect of network geometry on the convergence of SusP.
The underlying graph of the SK model is fully connected with many 
loops, so this is presumably a rather difficult case for SusP.
We therefore consider randomly diluted SK models, where a fraction $(1-c)$
of the couplings to each node are set to $0$
while $cN$ randomly picked couplings are drawn from a Gaussian distribution with variance $1/(cN)$
or $1/N$.
In the same spirit, but in the opposite direction, we also consider SusP
on a lattice graph, where each spin is connected to its $cN$ nearest neighbors.

Finally, we apply SusP to infer functional couplings in data generated from
a simulated neural networks. Using inverse models to describe the statistics of 
neural data has been one of the very active fields of research 
lately \cite{Schneidman05,Shlens06,Roudi09,CoccoLeiblerMonasson} and various approximations from statistical
mechanics have been studied for the inference of Ising model
applied to neural data \cite{Roudi09-2}, showing that TAP inversion and
Sessak-Monasson approximations work very well for this type of data. 
Here we show that for fine time bines ($~10$ms), SusP 
is outperformed by these simpler approximations.

To end this Section: for small smaples, \textit{i.e.} $N\le 20$,  we can calculate the means,
$m_i=N^{-1}\sum_{\bsigma} \Pr(\bsigma) \sigma_i$,
and correlations $C_{ij}=N^{-1}\sum_{\bsigma} \Pr(\bsigma) \sigma_i \sigma_j-m_i m_j$
by exact enumeration and summing over all the $2^N$ states. For larger values of $N$,
we use Monte Carlo sampling to estimate the means and correlations since the exact summation
is not feasible. For larger values of $N$ the means and correlations are therefore
always more or less noisy.

\section{Convergence at Low T}
\label{s:convergence}
In this Section we study convergence of the SusP algorithm on the
test case of an SK model at zero external field. Temperature,
as in Eq.\ \eqref{eq:SK}, should be understood as a shorthand of the
overall size of the couplings. It is well-known that this model
has a phase transition to a spin glass phase at $T_c=1$.  
When the temperature is high, the SusP algorithm can find the couplings
very accurately, with a reconstruction error that is essentially limited
by the quality of the measured means and correlation and the machine accuracy.
We do not show these data, already obtained by \cite{Marinari10},
but only point out that in the high temperature regime other reconstruction methods also work,
and importantly the high temperature TAP inversion technique would work well \cite{Kappen98,Tanaka98}.
Therefore, SusP can only be a competitive reconstruction method at high
temperature if both the quality criteria are very strict, and also the
means and correlations are known very accurately.

The interesting case is, therefore, at low temperatures, where most
reconstruction techniques have difficulties, and where the SK model approaches
the spin-glass phase.  
One way in which the procedure sketched above
and embodied by Eq.\ \eqref{eq:SusP} can definitely
fail to converge is 
if the argument of the right hand side of Eq.\ \eqref{eq:SusP} 
eventually becomes larger than one in absolute value. If so, inversion is
not possible (imaginary $J_{ij}$). 
A convenient empirical tool to monitor algorithm convergence and performance on 
families of large instances is a \textit{pivoting plot}. A number of random 
instances are generated and the convergence or performance of the algorithm is determined
across the family. The fraction of good outcomes then varies with the size of
the instances and a parameter describing the family. In the case at hand, this
parameter is the temperature $T$, and the size is $N$. In favorable cases
the location of the transition between high and low fractions of good outcomes 
depend only weakly on $N$ and becomes
sharper a $N$ grows, giving rise to a characteristic ``pivoting'' shape of the
cumulative empirical probability distributions. 

\begin{figure}[]
\centering
\includegraphics[height=6.35 cm, width=8.09 cm]{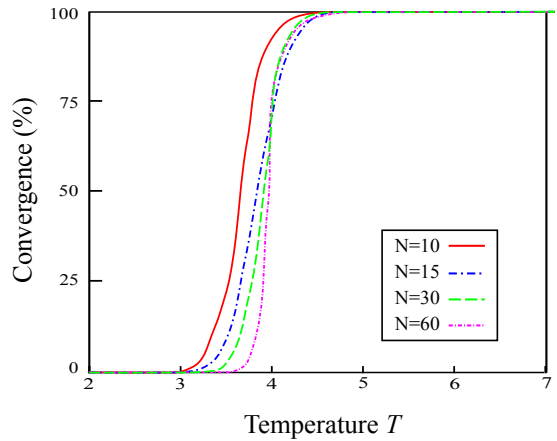}
\caption{Convergence of SusP with no stopping criterion or damping 
factor. A phase transition seems to occur around $T_{SusP}\approx 4$. We can conjecture 
that at higher $N$ the phase transition will be sharper.} \label{Fig1}
\end{figure}

As shown in Fig.\ \ref{Fig1}, for SusP on the SK model, there is clear pivoting
and a threshold separating a convergent from a non-convergent phase. This 
transition occurs at $T_{SusP}\approx 4$. To avoid this convergence problem, the update 
rule can be modified with a damping factor \cite{Marinari10}. 
There are two problems, however, with the low-T convergence even
when the damping factor is used. First, the damping factor
makes the convergence slow for low $T$ because strong damping $\epsilon$
must be employed. Second, we have also observed that for sufficiently low $T$, 
although the imaginary $J$ problem can be avoided by using the damping 
factor, the algorithm fails to converge even with very small $\epsilon$.

\begin{figure}[]
\centering
\includegraphics[height=12 cm, width=8.5 cm]{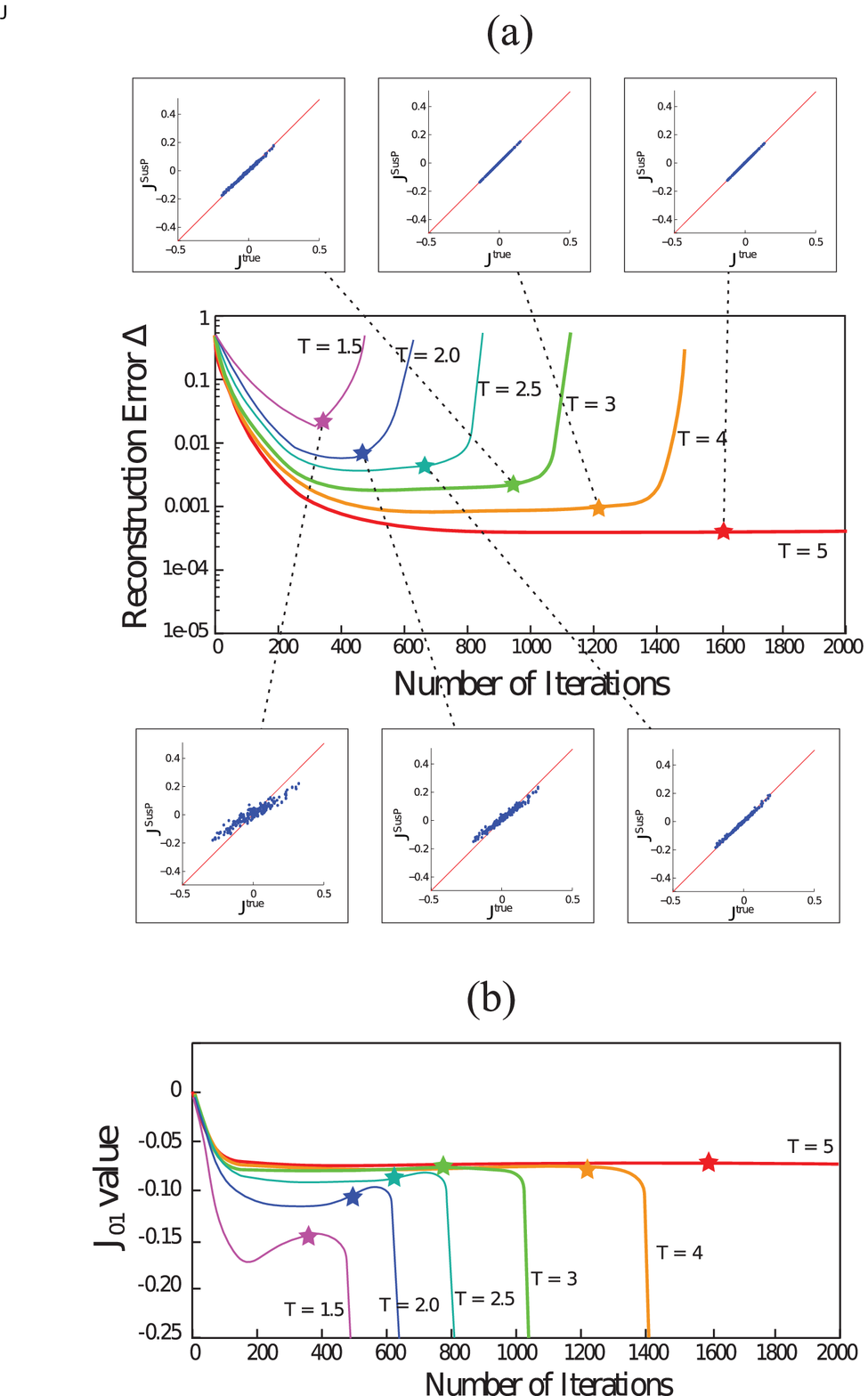}
\caption{The reconstruction error as the algorithm
proceeds (a), and the dynamics of one particular coupling $J_{01}$ (b)
as the algorithm proceeds, and for different temperatures.
The stars show where the stopping criterion Eq.\ \eqref{eq:stop}
is satisfied and the associated scatter plots show the 
inferred couplings versus those of the original model.} \label{Fig2}
\end{figure}

The behavior of the algorithm on representative instances are
exemplified in Fig.\ \ref{Fig2}, where 
we show total reconstruction error as well as the dynamics of one example coupling 
versus the number of iterations used in the algorithm. 
It is clearly seen that even for small $T$,
where the algorithm will eventually diverge, there is a plateau 
where the couplings remain almost constant for many iterations. 
In fact, comparing with the
dynamics of $J$ at larger $T$, where convergence does occur, 
we see that the dynamics is then qualitatively similar, only the plateau
then being also the asymptotically steady state. In other words, 
the dynamics of the SusP algorithm close to the solution
has a direction which changes from marginally stable to marginally
unstable as $T$ decreases, and this can be seen to be the
cause of bad reconstruction in this regime.
Consequently, 
by stopping the algorithm when changes in $J$ from one
iteration to the next are small, long before the argument of the
right hand side of Eq.\ \eqref{eq:SusP} turns negative,
we can hope to achieve good 
reconstruction also at very low temperatures. 

\begin{figure}[]
\centering
\includegraphics[height=7 cm, width=7 cm]{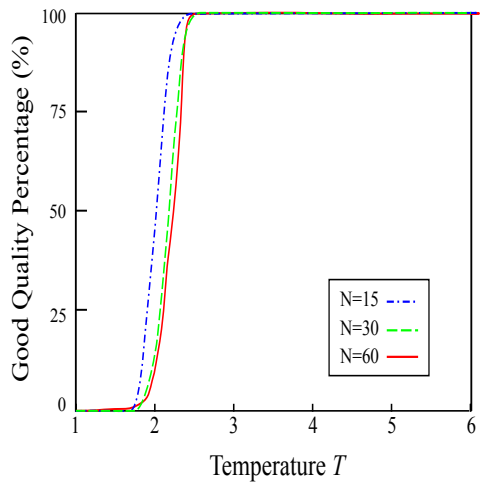}
\caption{Percentage of good reconstruction with stopping criterion. A reconstruction 
is considered good when the error $\Delta$ is below 0.05. There is a rather sharp 
threshold at temperature close to 2.} \label{Fig3}
\end{figure}

This is indeed possible, as shown in the scatter plots 
in Fig. \ref{Fig2} where we have shown the inferred 
couplings versus true couplings after $t$ iterations 
where

\begin{equation}
|J^{t}_{ij}-J^{t-1}_{ij}|>|J^{t-1}_{ij}-J^{t-2}_{ij}|
\label{eq:stop}
\end{equation}

\noindent for at least $90\%$ of the pairs of $i$ and $j$. 
Fig.\ \ref{Fig3}, shows how according to the criterion of 
at most 5\% reconstruction error (measured by $\Delta$) reconstruction
is possible at lower temperatures, down to new threshold $T_{SusP'}$
approximately equal to $2$.

\section{Influence of varying the external field and mean magnetization}
\label{s:magnetization}

The results shown in the previous section as well as those 
reported in \cite{Marinari10} were all run on data from a 
model without external field, \textit{i.e.} when $h_i=0$ for all
$i$. In this case, because of the symmetry of the problem 
the mean magnetizations are all zero (if calculated exactly), and fluctuate
around zero when computed by Monte Carlo sampling. This raises the question of 
the effect of magnetization and external fields on the reconstruction error.
 
As described in Appendix, SusP first computes the couplings in a
loop of alternate steps of changing $J_{ij}$ and
Belief Propagation updates; the external fields $h_i$ are only computed 
in a Belief Propagation output step. Similarly to other schemes such as TAP,
reconstruction of the external fields can therefore essentially never be better than 
the reconstruction of the couplings. The questions are therefore, first, how well 
the couplings are reconstructed if the magnetizations are sensibly different
from zero, and second, if reconstruction quality is degraded in the final step from
the couplings to the external fields. This is an important issue to investigate
since in many real life applications, \textit{e.g.} neural data binned at smal time
bins, one would be dealing with highly magnetized variables.

\begin{figure}[]
\begin{center}
\includegraphics[height=11 cm, width=7 cm]{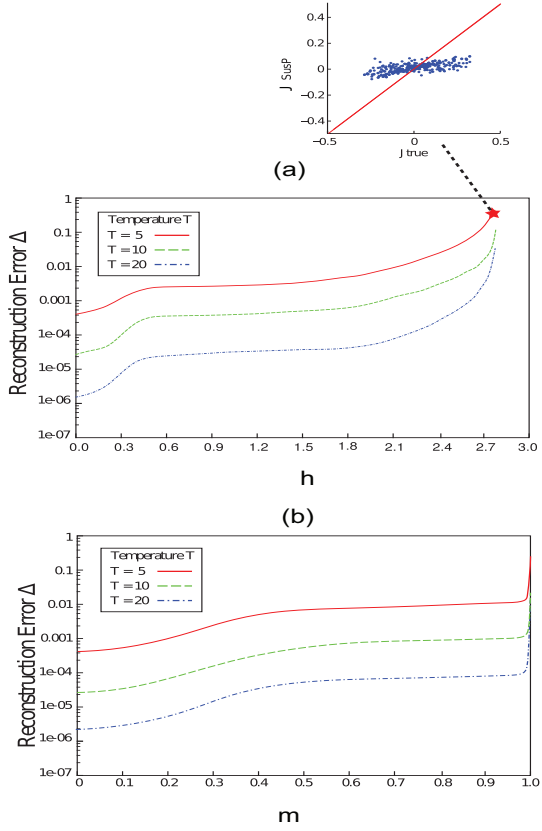}
\end{center}
\caption{Reconstruction error as a function of external field (a)
and mean magnetization (b). At all temperatures and network sizes tested, 
increasing the absolute magnitude of the magnetization 
by increasing the external field has a negative effect on the 
reconstruction.
} \label{Fig5}
\end{figure}

On the first point, in Fig. \ref{Fig5}, we show that reconstruction
of the couplings by SusP is degraded, either as a function of external fields, or as
function of mean magnetization. 
Here we have used a uniform external field applied to all spins.
As a function of external fields the loss of
reconstruction quality is smooth, while as a function of magnetization it
is abrupt, and concentrated in a narrow range close to $|m|\approx 1$, essentially due to the
fact that as $|m|\to 1$ one has $h=\tanh^{-1}(m)$ \cite{Roudi09}.
The dependence is sharper at higher temperature, \textit{i.e.} when SusP
in zero magnetization works well. 

\begin{figure}[]
\begin{center}
\includegraphics[height=6 cm, width=6 cm]{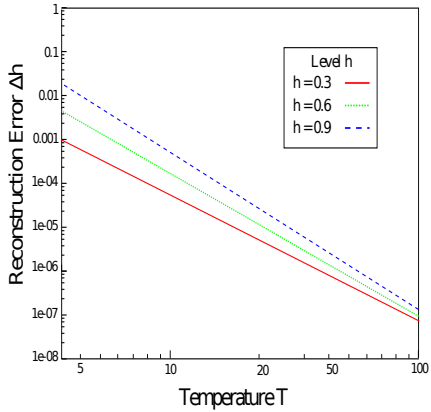}
\end{center}
\caption{The reconstruction error $\Delta h$ (mean value of the 
absolute error of each field) as a function of the temperature $T$.} \label{Fig4}
\end{figure}

On the second point, we do not observe significant degradation going
from the couplings to the external fields.
As shown in Fig. \ref{Fig4}, the quality of this 
reconstruction also degrades at lower tepmperatures, 
as is the case for the couplings.

\section{Geometry and sparsity of the graph}

SusP has its root in the Belief Propagation algorithm. It is well-known
that the BP is exact on trees, \textit{i.e.} graphs without 
loops \cite{YedidiaFreemanWeiss03}. We would
then expect that the performance of SusP is also influenced by
the presence or absence of loops in the model that generated the data.

\begin{figure}[]
\centering
\includegraphics[height=11 cm, width=8.5 cm]{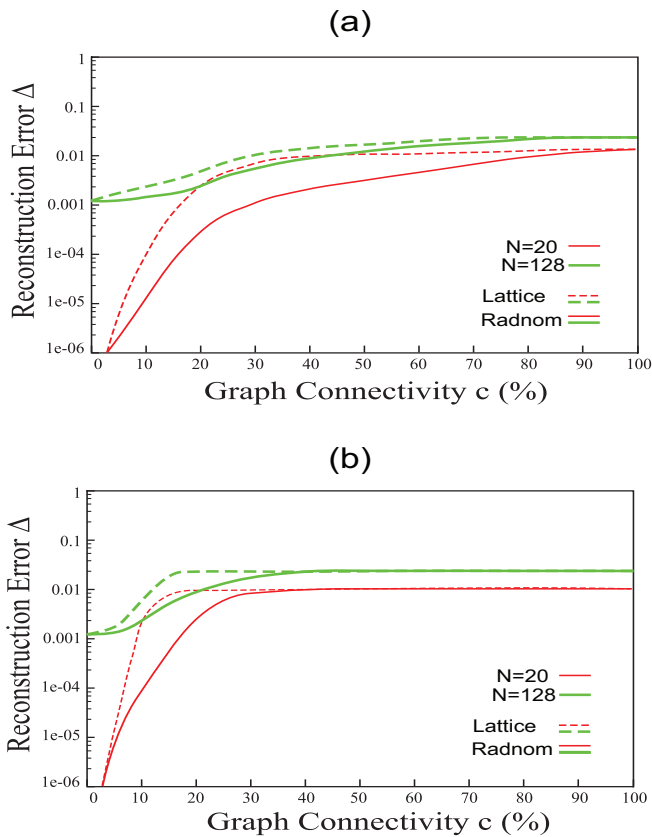}
\caption{The influence of the sparsity and
connectivity pattern on SusP. (left) The reconstruction error
is shown versus the fraction of connections for
different sizes both in the case of a random graph
and a 2D lattice. In this case
the variance of the couplings scale as $1/N$ independent of $c$. 
(right) This is the same as a, except that the variance of the
couplings now scale as $1/cN$.  SusP appears to work better 
on sparser and less loopy graph.} \label{Fig6}
\end{figure}

To test this hypothesis, we first studied how the sparsity of the
connectivity pattern of the model influences the reconstruction.
We thus generated data from a SK model in which a fraction $c$ of
the couplings are put to zero while the rest are drawn
either from a Gaussian distribution with variance $1/N$,
or with $1/(cN)$. The results are shown in Fig. \ref{Fig6}
where we plot the reconstruction error versus the sparsity $c$. 
We can observe a decrease in the reconstruction error as $c$ decreases,
thus concluding that SusP works better when the connectivity is sparse.
For almost all values of $c$, the positive effect of 
a sparser graph on the reconstruction error is suppressed when one has stronger
connections, \textit{i.e.} when the variance of the couplings are taken to be $1/(cN)$.

To understand how the graph geometry influences SusP, we altered 
the geometry of connections, at fixed $c$, from a random connectivity to a 2D 
rectangular lattice with each node being connected to $cN$ other closest nodes
to it. This way, at a fixed sparsity, we essentially increase the loopiness of
the graph, by inducing local small loops. As shown in Fig. \ref{Fig6}
the presence of these local loops increases the reconstruction error. For
a given connectivity, a random sparse graph has fewer local loops than the 
lattice one, which explains why the reconstruction is better on random sparse
than on a lattice.

\section{Tests on Neural Data}

The Ising model has been used in a number of studies for modeling the statistics of
muti-neuron spiking patterns of binned spike trains. They were shown to provide 
a good model for spike patterns over $N$ neurons, if the average number of spikes generated
by all $N$ neurons in a time bin is small \cite{Schneidman05,Roudi09}. Various approximations have been studied
in \cite{Roudi09-2,Roudi09-3} where TAP inversion and SM approximations were
shown to provide good estimates of the functional couplings. In what follows 
we study the performance of SusP on synthetic neural data generated form 
a simulated cortical network model and compare them 
with TAP. For details of the simulations see \cite{Roudi09-2}.

\begin{figure}[]
\centering
\includegraphics[height=9. cm, width=8.5 cm]{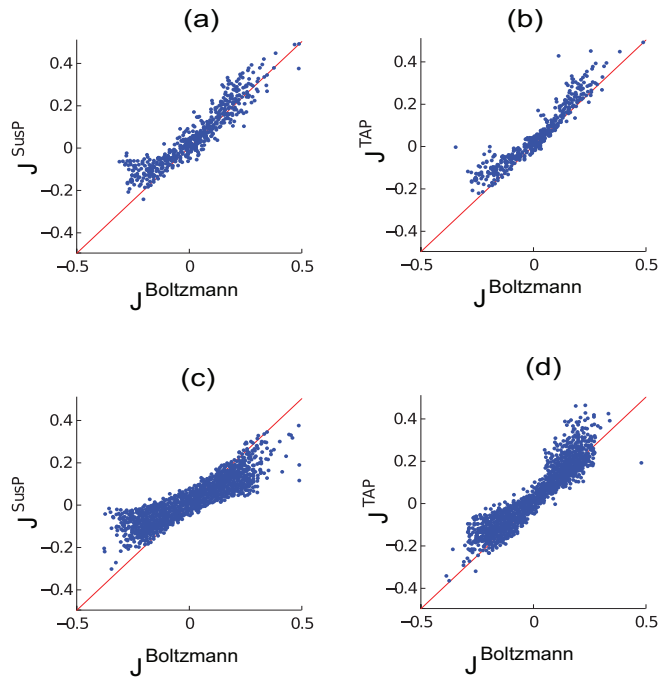}
\caption{Comparing SusP (left) and TAP results (right) versus the 
Boltzmann learning for $N=40$ (up) and $N=100$ (down) 
for synthetic neural data binned at $\delta t=10$ ms.} 
\label{Fig8}
\end{figure}

Fig.\ \ref{Fig8} shows the results of TAP, and SusP as compared to the 
results of Boltzmann learning for neural populations of size $N=40$
and $N=100$. This figure shows that for this type of data TAP
outperforms SusP, particularly for large $N$. One reason for this could be the fact that to make a binary representation
of the spike trains, one has to bin them to small time bins, typically of size $\delta t=10$ms. This yields
a mean probability of spike per bin of $\sim 0.9$ or mean magnetization of $\sim -0.7$. Our previous
analysis in section \ref{s:magnetization} did provide the conclusion that large absolute magnetization
has a negative effect on the performance of SusP. 

\section{Discussion}

The ability of experimentalists to observe the activity 
of a large number of elements in biological networks needs to be 
accompanied by mathematical tools to analyse the high dimensional
recorded data. One approach for doing this it to develop
analytical and numerical tools to infer a probability distribution
with a small number of parameters from the recorded data, small compared to the
possible number of states of the system. The fitted
model can then be used to generate synthetic data, or it could be 
used to learn something about the network that generated the data,
\textit{i.e.} to learn functional connections between the elements. 

The Ising model is probably too simple a model 
for many real life purposes and the inferred couplings may not
correspond to real physical interactions, depending on the real underlying
physical system \cite{Roudi09-2}. However, it provides an excellent
platform to study the analytical and theoretical aspects of 
the problem. Many of the hurdles encountered in the
inverse Ising problem are likely to be present when dealing with
more complicated models. Developing and analyzing approximate and iterative
methods for inverse Ising problems has therefore attracted a lot of 
attention in the past few years. 

The SusP algorithm is the result of the last effort in 
this direction. Not surprisingly, it exhibits complex dynamical behaviors
when applied to loopy graphs some of which we studied here. Importantly,
the simplest implementation of the algorithm exhibits a dynamic 
phase transition to a non-convergent regime at $T_{SusP}\approx 4$, i.e. higher than the
equilibrium phase transition of the SK model. Our numerical experiments 
show that for a range of temperatures 
below $T_{SusP}$ the lack of convergence of the algorithm is due to the presence of an
unstable direction in the trajectory of the couplings, such that the algorithm
gets close to the true couplings, stays there for a while and then moves away from it.
Exploiting this fact, it is possible to perform good reconstructions down to 
$T_{SusP'}\approx 2$. Whether it is possible to avoid such a direction all together requires
a more detailed study of the dynamics of the couplings during the
learning process. It would also be interesting to see whether loop correction
methods such as those developed for Belief Propagation \cite{Mooij07} 
could be also employed in SusP.

Two new observations of the SusP algorithm have been made in this
paper. One positive, albeit expected, is that SusP works much better (at
large typical couplings) on sparse graphs. If it is known a priori
that the underlying graph is sparse, then SusP should be very competitive 
to other schemes using means and correlations such as TAP or Sessak-Monasson.
However, this is also the regime where the recent $l_1$ reconstruction of 
Ravikumar, Wainwright and Lafferty~\cite{RavikumarWainwrightLafferty10}
provably works. More experimentation is therefore needed
to establish if SusP is competitive in this regime. Our second observation,
negative, is that reconstruction by SusP degrades in a strong external field.
Qualitatively speaking, a strong external bias overwhelms the correlations
since spins do not fluctuate much, and a finite amount of data on correlated
fluctuations contains less useful actionable information for the SusP reconstruction.
Since the high-field limit is relevant to neural data, this may limit the
applicability of SusP in this domain. In tests on one series of synthetic
neural data, we indeed found that SusP does not work better than simpler
schemes such as TAP.

SusP has already been extended and used for non-binary systems by 
Weigt et al \cite{WeigtPNAS2009}. These authors extended the SusP algorithm
to Potts-like variable and successfully inferred direct
physical interactions between amino acids in a family of
two-component signal transduction pathways in bacteria.
SusP has therefore proven its usefulness
on a problem of significant biological interest.
The convergence
properties of this extended SusP in a suitable class of random models
have however not been systematically investigated.

\appendix
\section{The Susceptibility Propagation Algorithm}

This appendix contains the update rules for the Susceptibility 
Propagation algorithm and their demonstration.
The rules are stated in Subsection~\ref{a:SusP-statement}, and 
are derived in Subsection~\ref{a:SusP-derivation}.
The relation to Boltzmann machines is described in Subsection~\ref{a:SusP-interpretation}.

\subsection{Susceptibility Propagation Equations}
\label{a:SusP-statement}
\noindent \textbf{Initialization}: The messages $h_{i \rightarrow j}$ and $v_{i \rightarrow j,k}$ 
are chosen at random while the other messages $u_{j \rightarrow i}$ and $g_{i \rightarrow j,k}$ 
are set to $0$. The couplings $J_{ij}$ are initially also set to $0$.\\\\

\noindent \textbf{Update rules}:\\

\begin{subequations}
\begin{align}
&h_{i \rightarrow j} \leftarrow \mbox{arctanh}\: m_i - u_{j \rightarrow i} \label{SusP-update-h}\\
&g_{i \rightarrow j,k} \leftarrow \sum_{l \in \partial i \setminus j} v_{l \rightarrow i,k} + \delta_{i,k} \label{SusP-update-g}\\
&C_{ij} \leftarrow \frac{\chi_{ij} - g_{i \rightarrow j,j} (1-m_i^2)}{g_{j \rightarrow i,j}} + m_im_j \label{SusP-update-C}\\
&\tanh J_{ij}\leftarrow \epsilon \frac{C_{ij} - \tanh h_{i \rightarrow j} \tanh h_{j \rightarrow i}}{1 - C_{ij} \tanh h_{i \rightarrow j} \tanh h_{j \rightarrow i}} + (1-\epsilon) \tanh J_{ij}\nonumber\\
&\label{SusP-update-J}\\
&\tanh u_{i \rightarrow j} \leftarrow \tanh J_{ij} \tanh h_{i \rightarrow j}\label{SusP-update-u}\\
&v_{i \rightarrow j,k} \leftarrow g_{i \rightarrow j,k} \tanh J_{ij} \frac{1 - \tanh^2 h_{i \rightarrow j}}{1 - \tanh^2 u_{i \rightarrow j}}\label{SusP-update-v}
\end{align}
\end{subequations}

\noindent \textbf{Output rules}:\\

The couplings $J_{ij}$ are derived in the update step.
The external fields are computed from the converged 
messages and the magnetizations as
\begin{equation}
\label{SusP-output-h}
h_i \leftarrow \mbox{arctanh} \: m_i - \sum_{j \in \partial i} u_{j \rightarrow i}
\end{equation}

\subsection{Deriving the SusP update rules}
\label{a:SusP-derivation}
As a starting point, we will use the canonical update equations of Belief Propagation applied to the pairwise Ising model:

\begin{equation}
p_{i \rightarrow j}(\sigma_i) \propto e^{-h_i \sigma_i}\prod_{f_k \in \partial i \setminus f_j} q_{k \rightarrow i} (\sigma_i) \nonumber
\end{equation} 
\begin{equation}
q_{j \rightarrow i}(\sigma_i) \propto \sum_{\sigma_j = \pm 1} e^{J_{ij} \sigma_i \sigma_j} p_{j \rightarrow i}(\sigma_j) \nonumber
\end{equation} 
In these two equation, $p_{i \rightarrow j}$ and $q_{j \rightarrow i}$ are the messages exchanged over spins $i$ and $j$, and $\partial i$ are the spins in the neighbourhood of $i$. Proportionality means up to
a normalization \textit{e.g.} $\sum_{\sigma_i}p_{i \rightarrow j}(\sigma_i)=1$
and $\sum_{\sigma_i}q_{j \rightarrow i}(\sigma_i)=1$. Finally, to retrieve the marginals we have the following equation:
\begin{equation}
p_i(\sigma_i) \propto e^{-h_i \sigma_i}\prod_{f_k \in \partial i} q_{k \rightarrow i} (\sigma_i)\nonumber
\end{equation} 
where proportionality again means up to a normalization.

For Ising spins it is convenient to work with the log-likelihood notation
\begin{equation}
h_{i \rightarrow j} = \frac{1}{2} \log \frac{p_{i \rightarrow j}(+1)}{p_{i \rightarrow j}(-1)}\nonumber
\end{equation}
\begin{equation}
u_{i \rightarrow j} = \frac{1}{2} \log \frac{q_{i \rightarrow j}(+1)}{q_{i \rightarrow j}(-1)}\nonumber
\end{equation}
which satisfy
\begin{equation}
h_{i \rightarrow j} = h_i + \sum_{k \in \partial i \setminus j} u_{k \rightarrow i} \nonumber
\end{equation}
and (in the other direction)
\begin{equation}
\tanh u_{i \rightarrow j} = \tanh J_{ji} \tanh h_{i \rightarrow j} \nonumber
\end{equation}
This is Eq.\ \eqref{SusP-update-u} of the SusP update rules.

\noindent SusP works also with the gradient of messages
\begin{equation}
g_{i \rightarrow j,k} = \frac{\partial h_{i \rightarrow j}}{\partial h_k},\,\,\,\,\,\,  v_{i \rightarrow j,k} = \frac{\partial u_{i \rightarrow j}}{\partial h_k} \nonumber
\end{equation}
Their update rules are derived by differentiating the update rules
of the messages with respect to external fields, and give
\begin{equation}
g_{i \rightarrow j,k} = \sum_{l \in \partial i \setminus j} v_{l \rightarrow i,k} + \delta_{i,k}\nonumber
\end{equation}
and
\begin{equation}
v_{i \rightarrow j,k} = g_{i \rightarrow j,k} \tanh J_{ij} \frac{1 - \tanh^2 h_{i \rightarrow j}}{1 - \tanh^2 u_{i \rightarrow j}}\nonumber
\end{equation}
These are Eqs.\ \eqref{SusP-update-g} and \eqref{SusP-update-v} of the SusP update rules.

The log-likelihood quantities describing the marginals
are the effective fields $H_i=\frac{1}{2} \log \frac{p_i(+1)}{p_i(-1)}$.
\textit{Exact} effective fields are related to magnetizations
$m_i$ and correlation functions $\chi_{ij}$ by
\begin{equation}
m_i = \tanh H_i \nonumber
\end{equation}
and (by fluctuation-dissipation)
\begin{equation}
\chi_{ij} = \frac{\partial \tanh H_i}{\partial h_j} \nonumber
\end{equation}
On the other hand, in Belief Propagation the effective fields
are related to the log-likelihood messages as
 \begin{equation}
H_i = h_i + \sum_{j \in \partial i} u_{j \rightarrow i}\nonumber
\end{equation}
Combining the two, or, what is the same, taking the Belief Propagation
computed effective fields as proxies for the exact effective fields,
we have  
 \begin{equation}
h_i = \tanh^{-1}(m_i) - \sum_{j \in \partial i} u_{j \rightarrow i}\nonumber
\end{equation}
which is Eq.\ \eqref{SusP-output-h}, the SusP output rule.
Substituting $h_i$ from above into the update equation for
$h_{i \rightarrow j}$ we have
\begin{equation}
h_{i\rightarrow j} = \tanh^{-1}(m_i) - u_{j \rightarrow i}\nonumber
\end{equation}
which is Eq.\ \eqref{SusP-update-h} of the SusP update rules.

If the correlation functions $\chi_{ij}$ are expressed in terms of the 
partial derivatives of the effective
fields, and the effective fields are computed by Belief Propagation, then
we will show that 
\begin{equation}
\label{eq:chi-SusP}
\chi_{ij} = \overline{\chi_{ij}} g_{j \rightarrow i,j} + g_{i \rightarrow j,j} (1 - m_i^2) 
\end{equation}
where the auxiliary quantities $\overline{\chi_{ij}}$ are
\begin{equation}
\label{eq:chi-SusP-aux}
\overline{\chi_{ij}} = \frac{\tanh J_{ji} + \tanh h_{i \rightarrow j} \tanh h_{j \rightarrow i}}{1 + \tanh J_{ji} \tanh h_{i \rightarrow j} \tanh h_{j \rightarrow i}} - m_i m_j 
\end{equation}
These equations do not (in general) hold exactly, but only under the assumption
that Belief Propagation is exact. If and when so, then by fluctuation-dissipation
\begin{eqnarray}
\chi_{ij} &&= \frac{\partial \tanh H_i}{\partial h_j} = \frac{\partial m_i}{\partial h_j}\nonumber \\
&&= \frac{\partial ( h_i + \sum_{j \in \partial i} u_{j \rightarrow i} )}{\partial h_j} (1 - m_i^2) \nonumber \\
&&= \frac{\partial ( h_{i \rightarrow j} + u_{j \rightarrow i} )}{\partial h_j} (1 - m_i^2) \nonumber \\
&&= (v_{j \rightarrow i,j} + g_{i \rightarrow j,j} ) (1 - m_i^2) \nonumber 
\end{eqnarray}
To show Eq.\ \eqref{eq:chi-SusP} and Eq.\ \eqref{eq:chi-SusP-aux}
we hence have to prove that $v_{j \rightarrow i,j}(1 - m_i^2) = \overline{\chi_{ij}} g_{j \rightarrow i,j}$,
which is the same as
\begin{equation}
\label{eq:main-equation}
\frac{1 - c^2}{1 - a^2c^2}(1 - (\frac{b + ac}{1 + abc})^2) = \frac{a + bc}{1 + abc} - \frac{c + ab}{1 + abc} \times \frac{b + ac}{1 + abc}
\end{equation}
using the shorthand
\begin{equation}
a = \tanh J_{ij} \quad b = \tanh h_{i \rightarrow j} \quad c = \tanh h_{j \rightarrow i} \nonumber
\end{equation} 
such that
\begin{equation}
m_i = \frac{b+ac}{1+abc}\quad m_j = \frac{c+ab}{1+abc}\quad \tanh u_{j \rightarrow i} = ac \nonumber
\end{equation}
Eq.\ \eqref{eq:main-equation} is an algebraic identity, as
can verified by termwise comparison, which therefore shows 
Eq.\ \eqref{eq:chi-SusP} and Eq.\ \eqref{eq:chi-SusP-aux}.  

In the final step we express $J_{ij}$ as a function of $\chi_{ij}$ and all the
other quantities.
It is convenient to first introduce yet another auxiliary quantity
\begin{equation}
\label{eq:C-def}
C_{ij} = \frac{\chi_{ij} - g_{i \rightarrow j,j} (1-m_i^2)}{g_{j \rightarrow i,j}} + m_i m_j
\end{equation}
which is Eq.\ \eqref{SusP-update-C} of the SusP update rules.
In Eq.\ \eqref{eq:C-def} the correlations $\chi$ are
from data. If, on the other hand, we substitute 
Eq.\ \eqref{eq:chi-SusP} and Eq.\ \eqref{eq:chi-SusP-aux} into Eq.\ \eqref{eq:C-def} we have 
\begin{equation}
\label{eq:C-another}
C_{ij} = \frac{\tanh J_{ji} + \tanh h_{i \rightarrow j} \tanh h_{j \rightarrow i}}{1 + \tanh J_{ji} \tanh h_{i \rightarrow j} \tanh h_{j \rightarrow i}} 
\end{equation}
which can be inverted to
\begin{equation}
\label{eq:tanhJ}
\tanh J_{ji} = \frac{C_{ij} - \tanh h_{i \rightarrow j} \tanh h_{j \rightarrow i}}{1 - C_{ij} \tanh h_{i \rightarrow j} \tanh h_{j \rightarrow i}}
\end{equation}
which is Eq.\ \eqref{SusP-update-J}, the final update rule for SusP,
and also the central step (Eq.\ \eqref{eq:SusP}) quoted in the
bulk of the paper. 

\subsection{Relation of SusP to Boltzmann machines}
\label{a:SusP-interpretation}

We recall that the Boltzmann machine approach consists of
the iterative procedure

\begin{equation}
\delta J_{ij} = \eta\left(\chi^{\rm Data}_{ij} -  \chi^{\rm Model}_{ij}\right)
\end{equation}

In practical applications this is often prohibitively expensive, because
the correlations $\chi^{\rm Model}_{ij}$ have to be evaluated by Monte Carlo.
Eq.\ \eqref{eq:chi-SusP} and Eq.\ \eqref{eq:chi-SusP-aux} are a means to quickly 
(if approximately) evaluate the correlations $\chi^{\hbox{Model}}_{ij}$ using
Belief Propagation variables and gradients of Belief Propagation variables.

The fixed point of a Boltzmann machine approach must satisfy the set of equations

\begin{equation}
\chi^{\rm Data}_{ij} = \chi^{\rm Model}_{ij}\left(\{J_{ij}\}\right)
\end{equation}

where we have made explicit that the correlations $\chi^{\rm Model}_{ij}$ depend
on all the model parameters $\{J_{ij}\}$. Eq.\ \eqref{eq:chi-SusP} is of just this
form, and so is Eq.\ \eqref{eq:C-another} for the auxiliary quantities $C_{ij}$.
A standard way to solve such equations would be by the Newton method or
other general-purpose root-finding methods.
Eq.\ \eqref{eq:tanhJ} however means to solve each equation for $\chi_{ij}$
separately by varying the corresponding $J_{ij}$, and keeping all else
fixed.  

Iterating Belief Propagation until convergence and then inverting locally
for $J_{ij}$ potentially wastes computation time in iterations. The 
actual scheme of \cite{MezardMora}
described above is therefore to interleave Belief Propagation update steps
(including updating the BP gradient quantities), and locally solving for $J_{ij}$.
This explains the order of the operations stated in \ref{a:SusP-statement}.
Convergence of this scheme is however not an easy question to answer. Presumably
there can be complications analogous to using Newton method
in a high-dimensional situation (one has to have a good starting point),
but as discussed in the bulk of the paper, there also seem to be other
effects (eigenvalue cross-over before the spin glass transition).  

\bibliographystyle{abbrv}
\bibliography{SusP}

\begin{thebibliography}{10}

\bibitem{Ackley85}
D.~H. Ackley, G.~E. Hinton, and T.~J. Sejnowski.
\newblock A learning algorithm for boltzmann machines.
\newblock {\em Cognitive Science}, 9:147--169, 1985.

\bibitem{CoccoLeiblerMonasson}
S.~Cocco, S.~Leibler, and R.~Monasson.
\newblock Neuronal couplings between retinal ganglion cells inferred by
  efficient inverse statistical physics methods.
\newblock {\em Proc Natl Acad Sci U S A}, 106:14058--62, 2009.

\bibitem{Kappen98}
H.~J. Kappen and F.~B. Rodriguez.
\newblock Efficient learning in boltzmann machines using linear response
  theory.
\newblock {\em Neur. Comp.}, 10:1137--1156, 1998.

\bibitem{Lezonetal06}
T.~R. Lezon, J.~R. Banavar, M.~Cieplak, A.~Maritan, and N.~Fedoroff.
\newblock Using the principle of entropy maximization to infer genetic
  interaction networks from gene expression patterns.
\newblock {\em Proc. Natl. Acad. Sci.}, 103, 2006.

\bibitem{Marinari10}
E.~Marinari and V.~V. Kerrebroeck.
\newblock Intrinsic limitations of the susceptibility propagation inverse
  inference for the mean field ising spin glass.
\newblock {\em J. Stat. Mech.}, 2010.

\bibitem{MezardMontanari09}
M.~M\'ezard and A.~Montanari.
\newblock {\em Information, Physics and Computation}.
\newblock Oxford University Press, 2009.

\bibitem{MezardMora}
M.~M\'ezard and T.~Mora.
\newblock Constraint satisfaction problems and neural networks: A statistical
  physics perspective.
\newblock {\em Journal of Physiology-Paris}, 103:107--113, 2009.

\bibitem{Mooij07}
J.~M. Mooij and H.~J. Kappen.
\newblock Loop corrections for approximate inference on factor graphs.
\newblock {\em Journal of Machine Learning Research}, 8:1113--1143, 2007.

\bibitem{OllionMScthesis}
C.~Ollion.
\newblock {Susceptibility Propagation for the inverse Ising Problem}, 2010.

\bibitem{RavikumarWainwrightLafferty10}
P.~Ravikumar, M.~Wainwright, and J.~D. Lafferty.
\newblock High-dimensional ising model selection using $l_1$-regularised
  logistic regression.
\newblock {\em Annals of Statistics}, 2010.
\newblock (to appear).

\bibitem{Roudi09-3}
Y.~Roudi, J.~A. Hertz, and E.~Aurell.
\newblock Statistical physics of pairwise probability models.
\newblock {\em Front. Comput. Neurosci.}, 3, 2009.

\bibitem{Roudi09}
Y.~Roudi, S.~Nirenberg, and L.~P. E.
\newblock Pairwise maximum entropy models for studying large biological
  systems: when they can work and when they can't.
\newblock {\em PLoS Comp. Biol.}, 5:e1000380, 2009.

\bibitem{Roudi09-2}
Y.~Roudi, J.~Tyrcha, and J.~Hertz.
\newblock Ising model for neural data: Model quality and approximate methods
  for extracting functional connectivity.
\newblock {\em Phys. Rev. E}, 79(5):051915, May 2009.

\bibitem{Schneidman05}
E.~Schneidman, M.~Berry, R.~Segev, and W.~Bialek.
\newblock Weak pairwise correlations imply strongly correlated network states
  in a neural population.
\newblock {\em Nature}, 440:1007--1012, 2006.

\bibitem{SessakMonasson}
V.~Sessak and R.~Monasson.
\newblock Small-correlation expansions for the inverse ising problem.
\newblock {\em J. Phys. A: Math. Theor.}, 42, 2009.

\bibitem{Sherrington75}
D.~Sherrington and S.~Kirkpatrick.
\newblock Solvable model of a spin-glass.
\newblock {\em Phys. Rev. Lett}, 35:1792 -- 1796, 1975.

\bibitem{Shlens06}
J.~Shlens, G.~Field, J.~Gauthier, M.~Grivich, D.~Petrusca, A.~Sher, A.~Litke,
  and E.~Chichilnisky.
\newblock The structure of multi-neuron firing patterns in primate retina.
\newblock {\em J. Neurosci.}, 26:8254--8266, 2006.

\bibitem{Tanaka98}
T.~Tanaka.
\newblock Mean-field theory of boltzmann machine learning.
\newblock {\em Phys. Rev. E}, 58:2302--2310, 1998.

\bibitem{WeigtPNAS2009}
M.~Weigt, R.~A. White, H.~Szurmant, J.~A. Hoch, and T.~Hwa.
\newblock Identification of direct residue contacts in protein-protein
  interaction by message passing.
\newblock {\em PNAS}, 106:67--72, 2009.

\bibitem{YedidiaFreemanWeiss03}
J.~S. Yedidia, W.~T. Freeman, and Y.~Weiss.
\newblock {\em Understanding Belief Propagation and its Generalizations}, pages
  239–--269.
\newblock Science \& Technology Books, 2003.

\end{thebibliography}
\end{document}